
\input phyzzx

\def\IGPP{\address{Institute for Geophysics and Planetary Physics,
      Lawrence Livermore National Laboratory, Livermore, CA 94550}}
\def\CFPA{Center for Particle Astrophysics, University
      of California, Berkeley, CA 94720}

\PHYSREV
\singlespace

\pubtype{CfPA-Th-91-010}
\titlepage
\singlespace
\title{Hopf Textures}
\author{Sun Hong Rhie$^{1,2}$ and David P. Bennett$^{2}$}
\address{${}^{1}$\CFPA}
\address{${}^{2}$\IGPP}

\abstract
\smallskip
A Hopf texture is a vacuum field configuration of isovector fields
which is an onto map from the space as a large three sphere
to the vacuum manifold $S^2$.
We construct a Hopf texture with spherically symmetric energy density
and discuss the topological charge. A Hopf texture collapses,
and we study the collapse process numerically.
In our simulations, it is clear that a Hopf texture does not decay
into a pair of monopoles. We also argue that the probability of forming
Hopf textures in random  processes is very small
compared to that of global monopoles.

\submit{Physical Review D}
\endpage


        \chapter{Introduction}

Field theories with spontaneously broken
symmetries have important roles in particle physics, early universe
cosmology, and condensed matter physics. Many of the most interesting
features of these theories involve topologically non-trivial field
configurations.
These can be classified by
continuous maps between spheres: $\pi_a (S^b): a\ge b$.
$\pi_1(S^1)$ describes cosmic strings
   \Ref\CamProc{G. W. Gibbons, S. W. Hawking, and T. Vachaspati, eds.,
      {\it The Formation and Evolution of Cosmic Strings}, Cambridge
      University Press, Cambridge, 1990.}
(local or global),
$\pi_2(S^2)$, describes magnetic monopoles (gauge) and global monopoles\rlap,
   \REFS\BV{M. Barriola, and A. Vilenkin, {\sl Phys. Rev. Lett.},
         {\bf 63}, 341, (1989).}
   \REFSCON\BR{D. P. Bennett and S. H. Rhie, {\sl Phys. Rev. Lett.},
        {\bf 67}, 1709 (1990).}
   \refsend
$\pi_3(S^3)$, describes instantons, baryon number violating anomalies
and global textures\rlap,
   \Ref\texture{F. Wilczek and A. Zee, {\sl Phys. Rev. Lett.}, {\bf 51},
    (1983) 2250;
    N. Turok, {\sl Phys. Rev. Lett.}, {\bf 63}, 2625, (1989).}
and finally $\pi_3(S^2)$ is the relevant algebra for Hopf
textures. The name Hopf texture refers to the name of the map
$\pi: S^3 \rightarrow S^2 (= CP^1) $, Hopf fibration.
(generally the Hopf map is $\pi: S^{2n+1} \rightarrow CP^n$)
\Ref\SPAN{ E. H. Spanier. {\it Algebraic Topology.} Springer-Verlag,
1966, pages 91 and 377.} The fibration induces the isomorphism:
\refmark\SPAN
$\pi_3(S^3) \rightarrow \pi_3(S^2)$ which in a sense is a sole reason
why $\pi_3(S^2)$ is nontrivial.
$\pi_4(S^3)=Z/2$,  `quantization' of two-flavour skyrmions,
\Ref\witten{E. Witten, {\sl Nucl. Phys.}, {\bf B223}, 433 (1983).}
$\pi_4(S^2)=Z_2$ and $\pi_4(S^4)=Z$ complete the nontrivial algebras
up to $a=4$, the dimension of conventional space-time.

In the standard big bang GUT cosmology the universe expands, cools
and goes through symmetry breaking phase transitions. At a GUT scale phase
transition, the formation of topological defects
\Ref\MONO {T. W. Kibble, {\sl J. Phys.} {\bf A 9} (1976) 1387.}
 such as cosmic
strings\rlap,\refmark\CamProc global monopoles\rlap,\refmark\BR or
textures can serve as seeds for the formation of galaxies and large
scale structure, and the dynamics of these objects is important for the
study of the cosmic structure they may seed. In this paper we will consider
the theories with symmetry breakings such as $O(3)\rightarrow O(2)$
which have two types of defects: global monopoles and Hopf textures.

When a global $O(3)$ is spontaneously broken to $O(2)$,
the vacuum states of the theory have a one to one correspondence with
the points on a two sphere.
Higgs fields prefer to align themselves (like ferromagnets)
within causally connected
volumes in order to minimize the gradient energy pressure.
Topological defects are believed to form in the
early universe because the coherence length of a field undergoing a
symmetry breaking phase transition must be small--smaller than the
causal horizon length. On scales larger than the initial horizon
length, the phases of the Higgs fields will be uncorrelated so that a
finite density of topological defects will be inevitable\rlap.\refmark{\MONO}
As the universe expands after the phase transition,
the horizon length grows and larger volumes come into causal contact.
This allows global monopoles to ``find" antimonopoles and annihilate
while Hopf textures can begin to feel the long range forces that cause
them to collapse.  As a consequence, the number of the
defects per horizon volume reaches a limiting value shortly after the
phase transition and remains constant. This is popularly referred to as
scaling behaviour, and it has been confirmed by numerical simulations
for cosmic strings\rlap,\refmark{\CamProc} global monopoles\rlap,\refmark{\BR}
domain walls\rlap,
   \Ref\PRS{W. H. Press, B. S. Ryden, and D. N. Spergel {\sl Astrophys.  J.}
        {\bf 357}, 293 (1990).}
and SO(4) textures\rlap.
   \Ref\STPR{D. N. Spergel, N. Turok, W. H. Press, and B. S. Ryden,
        {\sl Phys. Rev.} {\bf D43}, 1038 (1991).}

Textures are different from  other defects in that the
mapping between $S^3$ and $R^3$ is not as simple as the mapping between
$S^1$ or $S^2$ and $R^3$. The smaller spheres can be simply embedded
in $R^3$, but to make the correspondence between $S^3$ and $R^3$, one
has to identify a boundary surface in $R^3$ (typically at infinity)
with a point in $S^3$.
If we impose the boundary condition $\vec\phi =$ c constant at infinity and
consider space as a large three sphere, then a Hopf texture
is a vacuum field configuration of isovector fields whose values wrap
the vacuum manifold two sphere once and an $SO(4)$ (or $SU(2)$) texture is
a vaccum filed configuration of iso-four-vector fields whose values wrap
arround the vacuum manifold three sphere. However, the boundary condition
is not reasonable in the context of the early universe, because the
field will be uncorrelated on scales larger than the horizon.
Therefore the probability to form integer charge textures is very  small,
and the dominant form of the structures will be  `partial
textures' as was discussed by Borill {\it et al}
\Ref\ED {J. Borill, E. J. Copeland, and A. R. Liddle, {\sl Phys. Lett.}
                               {\bf B258}, (1991) 310.}
in the case of $SO(4)$ textures.  The notion of `partial texutre' makes
sense because the non-integer `partial texture' number is fairly well defined
in the given patch of space which contains high density of `fibers\rlap'.
This is due to that (as we shall see) the dynamical evolution
tends to intensify the texture density until the moment of collapse
when the texture number changes by 1.

Here we include a brief description of  $SO(4)$ textures which are simpler
to understand than Hopf textures.
A `partial $SO(4)$ texture' is a map from a patch of space
onto a patch of three sphere, and
the winding number is the fractional volume of the patch of three
sphere wrapped by the patch of space.
The dynamics tends to isolate the texture so that
at later times one can define a boundary surrounding the texture
that does have an approximately constant field value. For example,
lets consider the evolution of a partial texture of
total flux $Q (> 1/2)$ inside a finite box with fixed boundary conditions.
Such a configuration will evolve toward a configuration with a shrinking
$Q=1$ texture in the center and
a partial texture of charge $Q-1$ bound to the boundary surface.
If there were a repulsive core preventing the texture from collapsing,
the shrinking unit charge texture would be stabilized. An example of this
would be the thermal (random) production of Skyrmions with integer topological
charge ($=$ baryon number).

In the following, we employ a simple theory of self-interacting isovector
scalar fields and construct a Hopf texture with spherically symmetric
energy density as a transient configuration of the theory.
Using the notion of {\it fiber - equifield contour}, we develop a way to
understand the configuration particularly in terms of topological charge.
We observe the linkage of fibers which results from the twist of
transition functions ($SO(2)$ gauge fields) of the three sphere as a fiber
bundle (a Hopf bundle). To understand general configurations, we develop
algebraic expressions and then discuss the notion of ``partial winding"
of Hopf textures and the necessity of ``gauge invariance\rlap."
We note that the total energy of a Hopf texture is proportional to the
size of the texture
and study the process of collapse of Hopf textures by computer
simulations. We pay particular attention to collapse of a skewed Hopf texture
because of the claim
  \Ref\nlc{I. Chuang, R. Durrer, N. Turok, and B. Yurke, {\sl Science},
        {\bf 251}, 1336 (1991).}
that a skewed Hopf texture in nematic liquid crystal (NLC) decays into
a pair of a global monopoles and find that that is not the case in our
simulation. Of course, our system of isovector scalar fields is somewhat
different from that
of the NLC because the `equilibrium manifold' in NLC is $RP^2 = S^2/\pm$,
while the vacuum manifold in $SO(3)$ field theory is $S^2$.
The smallest integer charge Hopf texture in NLC wraps around the
`equilibrium manifold' twice. So we examined the evolution  of $Q=2$
Hopf texture of isovector scalar fields only to confirm the
conclusion with $Q=1$ case.

\chapter {Construction of a Hopf Texture}

Let's consider a theory of a self-interacting isovector scalar field
with a ``negative mass squared" term.
The Lagrangian is
$$ {\cal L} = {1\over 2} \partial_\mu\phi^a \partial^\mu\phi^a
   - {\lambda\over 4} (\phi^a\phi^a-\eta^2)^2 \quad , \eqn\lagrangian$$
and the equations of motion are
$$  \partial^2 \phi^a + \lambda (|\phi|^2-\eta^2)\phi^a=0\quad .
    \eqn\eqmotion $$
The vacuum states are given by  $|\phi| = \eta$, the vacuum expectation
value, and the set of the vacuum states is a two sphere.
Here $\eta$ is assumed to be $ \sim 10^{16}$ GeV, the generic GUT scale.
A global monopole is a static solution
$$ \phi^a = \eta f(r) {x^a \over r} \quad , \eqn\monopole $$
where $f(0)=0, f(\infty)=1$.
$f(r)$ rapidly approaches unity as $r$ becomes larger than the core size
$\delta = 1/\sqrt{\lambda\eta^2}$. When the scale of the physics of
interest is much larger than the core size, we can assume that $f(r)=1$
almost everywhere, and the $SO(3)$ sigma model
is a good approximation.
The eq.\eqmotion\ can be written as,
$$ \partial^2\phi^a
  = {\phi^a\phi^b\partial^2\phi^b \over|\phi|^2}
   +{\phi^a\partial^2|\phi|^2\over 2|\phi|^2}\quad ,\eqn\eqsigmaI $$
and in the limit $|\phi|=\eta$, we obtain
$$ \partial^2\phi^a-{1\over\eta^2}
        \phi^a \phi^b \partial^2\phi^b=0\quad .\eqn\eqsigmaII $$
$$  \phi^a\phi^a=\eta^2 \quad . \eqn\eqsigmaIII $$
These are the defining equations of the system of isovector scalar fields
we will discuss in this paper.

A global monopole has a singularity at the center and the field values
are the same along the radial lines stretching from the center to infinity.
If the center is removed, the punctured space $R^3 - \{*\}$
is homeomorphic to a `cylinder' $S^2 \times R$ and the trivial
fibration $\pi: S^2 \times R \rightarrow S^2$ amounts to a `winding'.
The number of `windings' is called the topological charge.
The triviality of the fibration makes the notion of fiber $R$
rather redundant in the analyses of global monopoles,
but when fields with two degrees of freedom are smoothly distributed
almost everywhere in three dimensional space, there are always one
dimensional curves along which the fields take the same values.
A {\it fiber} is nothing but an {\it equifield contour},
and a monopole has a singularity because fibers converge.
Of course, the region of the same field value can be a surface
or a patch of space (or even the whole space), but not an isolated
point. A fiber is either open or closed, and a finite open fiber ends on
a monopole.  A Hopf texture consists of closed fibers.

A Hopf texture is a mapping from space (a large ``three sphere")
onto the vacuum manifold two sphere.
The mapping is a nontrivial fibration
$\pi: S^3 \rightarrow S^2$, where the fiber is $S^1$.
In order to construct a Hopf texture, we have to find the projection map
$\pi$, and the first task is to find circles in $S^3$ (total space)
and contract each
of them to a point so that the points form $S^2$ (the base manifold).
We note  that $S^3$ is invariant under
$SO(4)$, and $SO(4)$ has $SO(2)$ subgroups which are circles and whose
modules are also circles. Thus we  parameterize $S^3$ such that
the circles in the three sphere are manifest.
For $(x_1,x_2,x_3,x_4) \in S^3 \subset R^4$,
$$ \eqalign{&x_1 = \cos(\xi) \cos(\eta),\cr
            &x_2 = \cos(\xi) \sin(\eta),\cr}
   \qquad
   \eqalign{&x_3 = \sin(\xi) \cos(\zeta),\cr
            &x_4 = \sin(\xi) \sin(\zeta),\cr} \eqn\eqxi  $$
where $0 \le \xi \le \pi/2, \ 0 \le \eta, \zeta \le 2 \pi$.
The volume element in this parameterization is
 $$ \sin(\xi) \cos(\xi) d\xi d\eta d\zeta  \quad . \eqn\eqvol $$

Consider an $SO(2)$ subgroup which rotates the 1-2 plane
and the 3-4 plane at the same time by $\eta \mapsto \eta - \psi$ and
$\zeta \mapsto \zeta - \psi$. Now project out the $SO(2)$ degree of
freedom by taking $\psi = \eta$,
where $\xi \not= 0$. Then $S^3$ is mapped to a hemisphere
$HS^2$, where the boundary $S^1$ corresponds to $\xi = 0$.
Contracting the boundary to a point by a smooth map completes the
Hopf fibration:
 $$ \eqalign{
   &\phi_1 = \sin(2\xi) \cos(\zeta-\eta) = 2(x_1 x_3 + x_2 x_4) \quad,\cr
   &\phi_2 = \sin(2\xi) \sin(\zeta-\eta) = 2(x_1 x_4 - x_2 x_3) \quad,\cr
   &\phi_3 = \cos(2\xi) = - 1 + 2(x_1^2 + x_2^2) \quad.\cr}
 \eqn\fieldI $$
Finally we obtain a Hopf texture with winding number unity
by mapping $S^3$ to $R^3_\infty (\equiv R^3 \cup \{\infty\})$.
We choose a stereographic projection which maps two spheres parallel
to the equator in $S^3$ to concentric spherical shells in three space.
$$ r(\chi) = D \tan ({\chi\over 2}) \quad , \eqn\rchi  $$
where $\chi$  is the polar angle of $S^3$ in spherical coordinates
$(\chi, \theta, \varphi)$.  $D$ is a scale parameter
being the radius of the spherical shell to which
the equator of three sphere is mapped.
The field distribution, with $D=1$, is
$$ \eqalign{
   &\phi_1 = {4y(1-r^2)+8xz\over (1+r^2)^2} \quad ,\cr
   &\phi_2 = {4x(1-r^2)-8yz\over (1+r^2)^2} \quad ,\cr
   &\phi_3 = -1+{8(x^2+y^2)\over (1+r^2)^2} \quad , \cr
                                      }  \eqn\fieldII  $$
and by replacing $\vec x$ by $\vec x / D$, one obtains the configuration
for an arbitrary $D$.

  \chapter {Topological Charge}

It should be  obvious from the construction that the field
configuration in eq. \fieldII\ wraps
the vacuum manifold once
and hence describes a texture of topological charge unity.
Here we develop an algebraic understanding of the topology of the
particular configuration in eq. \fieldII\ and extend it to arbitrary
configurations.

The charge density of $\pi_n (S^n)$ can be constructed using the
$\epsilon$-tensor which is antisymmetric and invariant under $SO(n)$.
For strings ($n=1$),
$$ Q = {1\over 2\pi} \int_{S^1} \epsilon_{ab} \phi_a d\phi_b  \quad , $$
  for  monopoles ($n=2$),
$$ Q = {1\over 8\pi} \int_{S^2} \epsilon_{abc}
        \phi_a d\phi_b d\phi_c \quad ,\eqn\chargeI $$
and for $SO(4)$ textures ($n=3$),
$$ Q = {1\over 12\pi^2} \int_{S^3} \epsilon_{abcd}
        \phi_a d\phi_b d\phi_c d\phi_d \quad ,\eqn\chargeII $$
where  three space is considered as a large three sphere $S^3$.
The simplest field configuration of a $SO(4)$ texture is the
homogeneous distribution of normal fields on $S^3$ ($\subset R^4$)
 stereographically projected onto $R^3$.
This is a simple {\it dimensional} extension of the
homogeneous normal field configuration on $S^2$ ($\subset R^3$)
which is just a cross section of the hedgehog monopole configuration
on a two sphere surrounding the pole.

In order to understand Hopf textures, we need a different line of
extension of the understanding of monopoles: {\it fibration}.
We interpret the eq. \chargeI\ for a monopole as:  one projects down
the fibers, {\it radial lines},
and see how many times the field configuration on the cross section
(here two sphere) covers the vacuum manifold two sphere.
If we apply the same idea to the Hopf texture in eq. \fieldII,
we should project down the fibers, {\it circles},
and see how many times the field configuration on the cross section
(say, `two sphere') covers the vacuum manifold two sphere.
Then the  charge formula should be
$$ Q = {1\over 8\pi}\int_{`two sphere'} \epsilon_{abc}
       \phi_a d \phi_b d \phi_c \quad , \eqn\chargeIII  $$
just as that of monopoles.
One remaining task is the identification of the `two sphere',
the submanifold dual to Hopf texture charge density two form.

Let's take a look at the field configuration in eq. \fieldII.
\FIG\FIELD {%
The explicit field configuration on the right show all the information but
the sign of $\phi_2$. The schematic diagram on the left presents the
extra information. The triplet of signs in each region represents the
sign of fields $(\phi_1, \phi_2, \phi_3)$ in the order.
}%
(see fig. \FIELD.)
The $z$-axis is a fiber corresponding to $\xi= {\pi\over 2}$ circle
and the field configuration is axially symmetric around it.
$\xi= 0 $ fiber lies on the $x-y$ plane having
the radius of $D\ (=1)$ and other fibers foliate a tube,
$ S^1 \times S^1$, stringed by $\xi= 0 $ fiber.
For example, the fibers of $\phi_3=0\ (\xi={\pi\over 4})$
loop around the donut shell
described by the equation $(\rho-\sqrt{2})^2+z^2=1$, where
$\rho^2=x^2+y^2$ so that each fiber winds once both of the $S^1$'s of
$S^1\times S^1$ (diagonal map from $S^1$ to $S^1 \times S^1$).
Therefore a fiber on the torus links the other fibers on the same torus
as well as both of the fibers $\xi=0$ and $\xi={\pi\over 2}$.
\FIG\TORUS {%
(a) Two diagonal non-intersecting curves on a donut shell
can be made  by  drawing parallel diagonal lines on a rectangle,
identifying the opposite edges and stretching the flat torus
to embed it in space.
(b) After one pair of edges are identified, the curves
are parallel spirals on a cylinder. If we fix the end points and
consider the curves as situated in three space, they are equivalent
to two open strings braided once. (c)  Identifying the other edges
closes the strings to loops which are linked.
}%
(See fig. \TORUS\ for the linkage of two diagonal non-intersecting
curves on a donut shell. Two non-diagonal non-intersecting closed curves
 on a torus have linking number zero.)
In fact, space can be thought of as a family of tori parameterized
by $\xi \in [0, \pi/2]$. In the limit of $\xi=0$, the radius of the
$left\ S^1$ converges to zero, $S^1 \times S^1 \rightarrow \{\ast\}
\times S^1 \approx S^1$ (the $\xi=0$ fiber), and in the limit of
$\xi=\pi/2$, the $right\ S^1$ collapses to a point,
$S^1 \times S^1 \rightarrow S^1 \times \{\ast\} \approx S^1$
(the $\xi=\pi/2$ fiber).
Note that the $left\ S^1$ is parameterized
by $\zeta$ which is ill-defined when $\xi=0$,
and $right\ S^1$ is parameterized
by $\zeta$ which is ill-defined when $\xi= {\pi\over 2}$ (see eq. \eqxi).
The linking of a fiber on an inner torus ($\xi$  smaller)
and a fiber on an outer torus ($\xi$  larger) should be
obvious because the fiber on the inner torus, as a
curve in three space, can be deformed to the circle $\xi=0$
without crossing the other fiber.
Thus, any fiber is linked with all the other fibers in the
Hopf texture with unit charge. In other words, if we choose any fiber,
the total flux will go through the loop of the fiber,
and the  `two sphere' in eq. \chargeIII\
 is an arbitrary surface bounded by the fiber.
For example, a half plane $\varphi=constant$, being bounded by the fiber
$\xi= {\pi\over 2}$, is a `two sphere'.
Indeed one obtains $Q=1$ by integrating the charge density over the
plane $\varphi=0$. Actually, this was to be expected from
the appearance of the hemisphere $HS^2$ bounded by $\xi=0$ in the course of
construction of the Hopf texture (below the eq. \eqvol). It should also
be obvious that the fiber $\xi=0$ is not any special than the others
because of spherical symmetry.

(A simpler example of `twist and linking' may be found
in  M\"obius strip, $\pi: M$\"o$ \rightarrow S^1$, where the fiber is
an interval $I=[0,1]$.
$s (\in I) = {1\over 2}$ is a circle, the M\"obius strip
is a family of circles parameterized by $s \in [{1\over 2},1]$,
and the circles link each other once.)

In summary, a surface subtending to the total flux of
a Hopf texture with unit charge is not a closed surface in three space.
It is an open surface bounded by a fiber, and we called it a `two sphere'
because its relative homotopy (the boundary is considered as one point,
because the field value is constant on it) is equal to $\pi_2(S^2)$,
whose algebra well known from  widely discussed monopoles.
Being derived from the same theory (the same fields and the same vacuum
manifold), the charge density of Hopf texture has the same functional form
as that of monopole, but Hopf textures are distinguishable from monopoles
because of the distinction in the submanifolds to integrate the charge
density over. For example, if we integrate Hopf charge density over a
closed surface, which is the proper submanifold to count monopoles over,
the integral vanishes. We can see this directly by integrating the charge
density of the field configuration in eq. \fieldII\ over the $y=0$ plane.
This is related to the fact that $\pi_2(S^3)=0$, namely the $y=0$ plane
can be deformed to infinity and contracted to a point.

Here we develop algebraic expressions to extend the intuitive
understanding above. In particular, we note that the fiber is
compact and the charge formula defined on a gauge slice
 can  be integrated along the fiber.
We will see the emergence of the connection, the gauge field.
Let's start with topological flux density
$$ {\cal B}_i = {1\over 8\pi} \epsilon_{ijk} \epsilon_{abc}
         \phi^a \partial_j \phi^b \partial_k \phi^c \quad ,  $$
which we can derive from the eq. \chargeIII.
The surface element corresponding to the flux density is
$$ dS^i = {1\over 2} \epsilon_{ijk} d{\rm x}^{jk} \quad ; \qquad
               d{\rm x}^{jk} \equiv d{\rm x}^i d{\rm x}^j  \quad . $$
We note that
$$ \nabla \times {\cal B} = 0 \quad , \eqn\norot $$
\ie, the flux is irrotational. If we compare it to magnetic fields,
it is easy to understand that that is due to the absence of currents
or line singularities.
\Ref\current {%
 In NLC, there are strings, $\pi_1 (RP^2) = \pi_1 (S^2/\pm)
  = \pm$, and $\nabla\times{\cal B}\neq 0$.
}%
However, there are point singularities (monopoles), and the
divergence does  not vanish at the singularities.
$$ \nabla\cdot{\cal B} = Q \, \delta^3 ({\rm x}-{\rm x}_0) \quad . $$
If there are no  monopoles,
the divergence vanishes everywhere, and this is the case we will
discuss in the following. For Hopf textures,
$$   \nabla\cdot{\cal B} = 0 \quad  $$
and there exists a vector potential $\cal A$ such that
$$    {\cal B} = \nabla \times {\cal A} \quad .  $$
The flux through an arbitrary surface $S$
$$ \int_S {\cal B} \cdot dS
        = \oint_{\partial S} {\cal A} \cdot d{\rm x} \quad , $$
and obviously the integral vanishes if the surface
$S$ is closed because $\partial S = 0$.
In cartesian coordinates of the
three sphere defined in eq. \eqxi, it's rather easy to find an ${\cal A}$
$$ {\cal B}_i = {1\over\pi} \epsilon_{ijk} (\partial_j x_1
      \partial_k x_2 + \partial_j x_3 \partial_k x_4 ) \quad , $$
$$ {\cal A}_i = {1\over 2\pi} (x_1 \partial_i x_2 - x_2 \partial_i x_1)
    + {1\over 2\pi} (x_3 \partial_i x_4 - x_4 \partial_i x_3)
                    \quad .  \eqn\eqA $$
\noindent
One can see that ${\cal A}$ is an $SO(2)$ gauge field
related to the simultaneous local rotation of the phases in the 1-2 plane
and the 3-4 plane. Under the rotation,
 ${\cal A} \rightarrow {\cal A} - \nabla \psi$,
where $\psi$ is the  rotation angle.
In spherical coordinates,
$$ {\cal B}_i = {1\over 2\pi} \epsilon_{ijk} \sin 2\xi\ \partial_j \xi
           + \sin^2\xi\ \partial_k \zeta) \quad ,  \eqn\aone $$
where the class $[\zeta - \eta]$ runs from $0$ to $2 \pi$.
In particular,
$$ \oint_{fiber} {\cal A}\cdot d{\rm x} = 1   \eqn\arot $$
because $\zeta - \eta = {\rm constant}$ along the fiber.
This is consistent with the fact that a `two sphere' is
bounded by a fiber for a Hopf texture with charge unity.
$$ 1 =  \int_{`two\ sphere'}  {\cal B} \cdot dS
     = \oint_{fiber} {\cal A}\cdot d{\rm x}    \quad .  \eqn\arotb $$
(Of course, the eq. \arot\ is not true in general as we will
demonstrate later with a field configuration of charge 2.)
Now we can change the charge formula in eq. \chargeIII\ into a volume
integral by multiplying by 1.
$$ \eqalign{ Q &=  \int_{gauge slice} {\cal B}\cdot dS
                    \oint_{fiber} {\cal A}\cdot d{\rm x} \cr
               &=  {1\over 2} \int {\cal A}_i
                       {\cal B}_j\  \epsilon_{jmn}\ d{\rm x}^{imn}  \cr
               &=  \int {\cal A}\cdot{\cal B}\ d^3{\rm x}  \quad . \cr
           }  \eqn\adotb $$
The second equality holds because ${\cal B}$ is the direction vector
of the fiber.
$$ \nabla_{{\cal B}} \vec\phi
    = {\cal B} \cdot \nabla \vec\phi = 0 \quad , $$
\ie, ${\cal B}$ is non-zero only when $\nabla\phi = 0$, which is
the defining equation of fiber.
The `differential charge' is the fractional volume of the three
sphere covered by $d^3{\rm x}$
$$  \delta Q = {\cal A}\cdot{\cal B}\ d^3{\rm x}
       = {1\over 2\pi^2} \sin\xi \cos\xi d\xi d\eta d\zeta \quad . $$
This displays the fact that the Hopf texture charge $Q$ arises because
three space $R^3$ is relatively a  three sphere here.
Of course, the fractional volume  does not define a (partial)
Hopf texture charge
because it is not a well-defined number. It is gauge variant. If we let
$$ \Delta Q = \int_{\Delta} {\cal A}\cdot{\cal B}\  d^3{\rm x} \quad , $$
then
$$ \Delta Q \rightarrow \Delta Q - \oint \psi{\cal B} \cdot dS \quad  $$
under a gauge transformation. This is not invariant, because $\psi$ has
nontrivial ($\neq$ constant) solutions.
For example, in a gauge  $\nabla\cdot {\cal A} = 0$,
$\nabla^2 \psi =0$, and $\psi=x$ is a solution.
Well, that's because $\Delta Q$  measures the winding of the
three sphere covered by the space patch $\Delta$, whereas a meaningful
measure  will have to be the winding of the vacuum manifold $S^2$.
In other words, $\Delta Q$ has to be associated with $S^2$ as well
as with the total space  $S^3$. The remedy is rather obvious:
the integral must include integration over the fiber.
If we rewrite eq. \adotb\ as
$$  Q = \oint_{fiber} {\cal A}\cdot d{\rm x}
       \oint_{\partial S} {\cal A}\cdot d{\rm x} \quad ,  \eqn\pureA $$
we see that a $fiber$ and $\partial S$ are the generators of a torus.
This is one of the family of tori we discussed before.
We also can write it as
$$ Q = \int_{S_f} {\cal B}\cdot dS \int_S {\cal B}\cdot dS \quad , $$
where $S_f$ is  a surface bounded by a fiber,
and we see that linking is essential because the first integral
would vanish if there were no linking.
For example, one can consider a vortex-like flux, say
${\cal B} = \hat\varphi /\rho^2$, then the flux is orthogonal to the plane
bounded by a fiber, and the integral vanishes.
(or ${\cal A} = \hat z / \rho$ and ${\cal A}\cdot{\cal B} = 0$.)
A Hopf texture charge is a count of linked fibers.

\section {Multiply Charged Hopf Textures}

Here we would like to discuss a couple of multiple integer charge
configurations. They are very minimal extensions of the case of $Q = 1$,
but they should provide a taste of general configurations.
Let's first consider a non-trivial field configuration of $Q = 0$
given by a map $\phi^{\prime}$. (It is understood that the spherical
coordinates of $S^3$ are functions on $R^3$ as given by the inverse of
the stereographic projection  in eq. \rchi.)
$$ \eqalign{
   &\phi_1^{\prime} = \sin(4\xi) \cos(\zeta-\eta) = 2 \phi_3 \phi_1 \cr
   &\phi_2^{\prime} = \sin(4\xi) \sin(\zeta-\eta) = 2 \phi_3 \phi_2 \cr
   &\phi_3^{\prime} = \cos(4\xi) = 2 \phi_3^2 - 1   \cr}
                 \eqn\fieldQzero  $$
$$ {\cal B}_i = {1\over \pi} \epsilon_{ijk} \sin 4\xi\ \partial_j \xi\
              \partial_k [\zeta - \eta]  \quad ,  $$
$$ {\cal A}_k = {1\over 2\pi} (\cos^2 2\xi\ \partial_k \eta
           + \sin^2 2\xi\ \partial_k \zeta)  \quad .  $$
The total charge is zero, because the
vacuum manifold is wrapped twice in the opposite senses.
For example, if we integrate over $y = 0$ half-plane,
$$ \int {\cal B} = 0 \quad . $$
However,
$$  \oint_{fiber} {\cal A} = 1 \quad ,  \eqn\arotII $$
and according to Stokes' theorem, the fiber can not be the boundary
of the  cross section of the total flux.
As before, space is considered as a family of tori parameterized
by $\xi$, and  the whole torus of  $(\rho - {\sqrt 2})^2 + z^2 = 1$
is a `fiber' with $\phi_3 = 1$.
The total flux inside the torus is $1$, and that outside of it is $-1$.
The fibers of the same field value inside and outside the torus
are distinct loops, and they link each other.
There is no oriented surface bounded by a pair of linked loops.
Now we consider a field configuration of $Q = 2$.
$$ \eqalign{
   &\phi_1^{\prime\prime} = \sin(2\xi) \cos 2(\zeta-\eta)
                    = {\phi_1^2 - \phi_2^2
               \over \sqrt{\phi_1^2 + \phi_2^2}} \cr
   &\phi_2^{\prime\prime} = \sin(2\xi) \sin 2(\zeta-\eta)
                    = {2 \phi_1 \phi_2
               \over \sqrt{\phi_1^2 + \phi_2^2}}  \cr
   &\phi_3^{\prime\prime} = \cos(2\xi) = \phi_3  \quad , \cr}
 \eqn\fieldQII $$
Here each (non-degenerate) torus has two fibers with
the same field values: for example,
$\zeta = \eta$ and $\zeta = \eta + \pi$.
A gauge slice, \eg,  $y=0$ half plane, cut through both of them in the
same sense, and the flux density shows the double strength.
$$ {\cal B}_i = {1\over 2\pi} \epsilon_{ijk} \sin 2\xi\ \partial_j \xi
              \partial_k (\zeta - \eta) \ , $$
and
$$ {\cal A}_k  = {1\over \pi} (\cos^2\xi\ \partial_k \eta
                  + \sin^2\xi\ \partial_k \zeta) \ .  $$
If we integrate over a fiber,
$$ \oint_{fiber} {\cal A} =  2 \ , $$
and the construction of the volume integral formula for the charge
can not be given by \adotb.
Let's define the ``linking number" of a Hopf texture as:
the ``linking number" is one if a fiber with one field value links with
another fiber with another field value, the ``linking number" is two if
a fiber with one field value links with other two fibers with another
field value, the ``linking number" is four if two fibers with the same
field value link with other two fibers with another field value, \etc.
Then one can see that the field configuration in eq. \fieldQII\ has
a ``linking number" of four, and the configuration \fieldQzero\ has
a ``linking number" zero because any fiber is linked with two fibers
with the same value and the opposites senses. One also can
see that the eq. \adotb\ delivers the correct ``linking number" for
integer charge Hopf textures.

\section {The Scarcity of Partial Hopf Textures}

Now let's consider partial Hopf textures. We can construct one in the
following manner:
take a $Q=1$ configuration, and consider a cube centered on the center
of the Hopf texture. Imagine that this cube ``cuts" the fibers that it crosses,
so that there is a maximum radius torus $T_{\rm max}$ which is intact. Next,
extend the field values on the walls of the cube radially to infinity.
Then the total flux is given by the flux within $T_{\rm max}$, but the
``linking number" of the partial torus is one according to the
definition above because the ``linking number" does not care about
the total number of the fibers involved in the linking. The ``linking
number" cares only about how many ``identical fibers" are involved.
In fact, if we take additional account of  the linking of the loops
on the tori with infinite fibers, the total flux of fibers involved
is the same as that of $Q=1$. In other words, for any partial Hopf
texture, there is a counterpart integer Hopf texture.
Now recall that
$$  1 = \oint_{fiber} {\cal A} = \int_{S_f} {\cal B} \ . $$
For example, the topological flux through the fiber $\xi=0$ (the last
`torus' to be cut) has to be one. Generally
$$ \oint_{fiber} {\cal A} = n \qquad ; \quad n \in {\rm integer}
   \ . \eqn\ainteger $$
In other words, a partial Hopf texture
cannot be an isolated donut of linked fibers of the total flux
of an arbitrary real number. The donut has to be linked by topological
flux of another partial Hopf texture or infinite fibers such that
the condition  \ainteger\ is met.
The implication is that the probability to form a partial Hopf texture
is comparable to the probability to form a full charge Hopf texture,
and so the number density of any Hopf texture in the early universe
is expected to be  very small.
This is very different from the case of $SO(4)$ textures.
(Simplistically speaking, Hopf textures are more global than
$SO(4)$ textures.) We can make a crude estimate of the density of
Hopf textures per horizon volume by doing this calculation in a toy
universe in which the field takes only certain discrete values at points
on a cubic grid in space. (The field is assumed to vary smoothly between
the grid points.) This yields an estimate of about $10^{-4}$ Hopf textures
per horizon volume.

In general, the flux of Hopf extures, if any, coexist with that of global
monopoles, and the total flux of the fibers through an arbitrary surface
can not expressed as a loop integral over the boundary of the surface.
That is because a gauge fields corresponding to monopole flux is
singular, or  the integral of the flux over a surface bounded
by a given loop jumps as the surface passes the poles.
$$ \int_S {\cal B}\cdot dS \neq
    \oint_{\partial S} {\cal A}\cdot d{\rm x} \ .  $$
For example,  we can consider a configuration where
the flux of a pair of monopole and
antimonopole is wedged along the $z$-axis through the unit
charge Hopf texture.
The vector potential ${\cal A}$ along a closed fiber must be the same
as that of the unit charge Hopf texture,
and its loop integration must be equal to 1. However, the total
flux going through a surface bounded by a fiber is two if
the monopole flux goes through the surface.
Thus, counting Hopf textures, \eg, in a simulation, is rather hopeless
because of its global nature. The volume integral \adotb\ doesn't really
help for identifying the flux either, because ${\cal A}$ field is not a
functional of the primary fields $\vec\phi$.

\chapter{Numerical Study of Collapse}

The energy density of the Hopf texture configuration given in eq. \fieldI\ is
$$ {\cal E} = {16 D^2\over (D^2+r^2)^2} \quad .
\eqn\endens $$
The total energy $E=16\pi^2 D$ is proportional to the
scale  parameter $D$ which implies that the texture is unstable to collapse.
In the non-linear sigma model ($\lambda \rightarrow \infty$ limit),
the Hopf texture will shrink to a point.
With a more realistic  potential, when $D$ gets
sufficiently small, the energy density will be large enough so that the
field can climb over the potential barrier and the Hopf texture can unwind.
The initial gradient energy will be released mainly as goldstone boson
radiation
with a small amount of massive Higgs radiation coming from the last stages of
the collapse.

Even though the energy density, \endens\ is spherically symmetric, the
field configuration, \fieldI\ is
only axially symmetric. Thus the process of collapse can not
be spherically symmetric, and the  family of configurations parameterized
by the size $D$ in eq. \fieldII\ are not dynamically related.
\Ref\sphron{It is interesting to note that a sphaleron
has axially symmetric fields
and spherically symmetric energy density. However, a sphaleron is a
static solution even though unstable, while there is no static Hopf texture.
Axial symmetry of a sphaleron is in the vector fields differently from
that of Hopf textures which are scalar fields. Higgs field of the standard
sphaleron is spherically symmetric.
See N. S. Manton, {\sl Phys. Rev.} {\bf D28}, 2019 (1983).}

Because of the lack of symmetry, it is probably fruitless to
look for an analytic solution to the collapse of a Hopf Texture.
Here we study the collapse numerically by evolving the eq. \eqmotion\
with the constraints that $\vec\phi^2 \equiv 1$ and
$\dot{\vec\phi} \cdot\vec\phi = 0$.
Unlike its continuum counterpart,
the numerical ``nonlinear sigma-model" does not force
$\vec\phi^2 = 1$ except at the locations of the grid points. Thus, objects such
as global monopoles or unwinding textures can be present in the numerical
``nonlinear sigma-model" simulations. In fact, it is not difficult to show that
the numerical sigma model is indeed the limit of the numerical version of
\eqmotion\ when $\lambda \rightarrow\infty$. The advantage of the numerical
``nonlinear sigma-model" is that one does not need to use extremely small
timesteps to prevent the instabilities in the massive degree of freedom
that occur when $\lambda$ is large. (The timesteps do have to be much
smaller than the spacial steps in order to evolve short wavelength modes
properly, however.) The results displayed below have used timesteps of
$\Delta t = 0.032 \Delta x$ for the ``nonlinear sigma-model" and
$\Delta t = 0.08 \Delta x$ for the ``$\lambda \phi^4$" simulation.
These parameters always resulted in global energy conservation of better than
$0.1\,$\% (typically $\sim 0.01\,$\%).

In our numerical simulations, we chose an initial configuration of unit
charge Hopf texture whose total flux is contained in a finite volume.
We chose the following map from $S^3 \rightarrow R^3$,
in place of the eq. \rchi ,
$$ r(\chi) = {R\over \pi} \chi \ , $$
and ran the following initial field configuration.
$$ \eqalign{\phi_1 &= 2\sin^2\left({\pi r\over R}\right) {xz\over r^2}
                    + \sin\left({2\pi r\over R}\right) {y\over r} \ , \cr
            \phi_2 &= 2\sin^2\left({\pi r\over R}\right) {yz\over r^2}
                    - \sin\left({2\pi r\over R}\right) {x\over r} \ , \cr
            \phi_3 &= 1 - 2\sin^2\left({\pi r\over R}\right) {x^2+y^2\over r^2}
                    \ , \cr}
\eqn\symfield $$
where $R$ is the radius of the Hopf texture.
\FIG\sympict{This figure shows a time series of the $y=0$ cross section of a
collapsing axially symmetric Hopf texture as computed in a $96^3$
``nonlinear sigma-model" simulation. Only $\phi_1$ and $\phi_3$
are shown which means that only the sign of $\phi_2$ is not determined
by the direction and length of the vector shown. (a) shows
the full cross section of the initial configuration
given by eq. \symfield\ with only every fourth grid point plotted.
(b)--(d) show the central quarter of the $y=0$ cross section at times
$t =  25.6$, 28.8, and $35.2 \Delta x$ respectively. In (b)--(d), every point
in the central region is shown.}
A time series of the collapse of this configuration is shown in Fig. \sympict.
This figure shows the $y=0$ cross section of eq. \symfield. Note that the
sign of $\phi_3$ cannot be determined from the information in this figure,
but that these configurations have a $\phi_3(-x)=\phi_3(x)$ reflection
symmetry (in the $y=0$ plane). Fig. \sympict (a) shows the $y=0$
cross section of the initial configuration with only 1 out of every 4
points shown for clarity. Figs. \sympict (b) and (c) show the
same cross section (magnified by a factor of 2) just before and just after
the decay of the Hopf texture. Fig. \sympict (d) shows this the field
configuration about 8$\Delta x$ after the collapse.
Note the similarity of the configurations long before (Fig. \sympict (a))
and very shortly before the collapse (Fig. \sympict (b)).

We have examined our simulations of Hopf texture collapse with a
monopole detection algorithm in order to test the claim by Chuang,
\etal \refmark\nlc
that Hopf textures in uniaxial nematic liquid crystals decay into
global monopoles.
This algorithm was designed to work with
the ``nonlinear sigma-model" simulations, and it detects monopoles in
the grid cells by smoothing interpolating the field from the verticies to the
sides of the cells.
For the simulations with axially symmetric initial conditions,
we found that the monopole detection algorithm never detected any monopoles.

\FIG\symcoll{The time evolution of various fibers intersecting the $y=0$
plane is shown at time intervals of $3.2\Delta x$ from $t=0$ to Hopf
texture collapse at $t\simeq 27\Delta x$. The solid lines and filled circles
represent the evolution of the $\phi_3=-1$ fiber, while the other curves
represent fibers in the $\vec\phi = (\cos\alpha,\sin\alpha,0)$ family
where $\alpha$ is an odd multiple of $\pi/8$. Each fiber is represented
by a different symbol. (Since each fiber is a loop on a torus,
they each intersect the $y=0$ plane twice.)}

Fig. \symcoll\ shows the flow of a selected set of fibers during the collapse
of the symmetric Hopf texture. After an initial transient, each fiber
tends to move at a nearly constant velocity until collapse, but these
velocities are not radial. So as expected, there seems to be no self-similar
collapse mode as there is for $SO(4)$ textures.

In order to understand how a more generic Hopf texture might collapse we
have also studied the collapse of an asymmetric Hopf texture with an
initial configuration given by,
$$ \eqalign{\phi_1 &= 2\sin^2\left({\pi r\over R_\phi}\right) {xz\over r^2}
                    + \sin\left({2\pi r\over R_\phi}\right) {y\over r} \ , \cr
            \phi_2 &= 2\sin^2\left({\pi r\over R_\phi}\right) {yz\over r^2}
                    - \sin\left({2\pi r\over R_\phi}\right) {x\over r} \ , \cr
            \phi_3 &= 1 - 2\sin^2\left({\pi r\over R_\phi}\right)
                    {x^2+y^2\over r^2} \ , \cr}
\eqn\skewfield $$
where
$$ R_\phi \equiv {R\over 3} \left( 2 + {x\over\sqrt{x^2+y^2}} \right) \ .
\eqn\Rphi $$
\FIG\skewpict{This figure shows a time series of the $z=0$ cross section of a
collapsing asymmetric Hopf texture as computed in a $96^3$
``nonlinear sigma-model" simulation. $\phi_1$ and $\phi_2$
are shown, so the sign of $\phi_3$ is not determined
by the direction and length of the vector shown. (a) shows
the cross section of the initial configuration
given by eq. \skewfield\ with only every fourth grid point plotted.
(b)--(f) show the central quarter of the $y=0$ cross section at times
$t =  12.8$, 14.4, 15.0, 16.6, and $19.8 \Delta x$ respectively.
Note that the figures (b)--(f) are magnified by a factor of $\sim 1.36$ with
respect to figure (a).}
Figure \skewpict\ shows a cross section of a time series of a collapsing
asymmetric Hopf texture. Note that the cross section shown is orthogonal
to the cross section shown in Fig. \sympict. For the symmetric collapse, this
cross section (the $z=0$) plane would display the axial symmetry.

Fig. \skewpict (a) shows the initial configuration while Fig.
\skewpict (b) shows the configuration immediately before the texture has
Hopf texture has decayed.  Figs. \skewpict (c) at $t=15.0\Delta x$
shows the configuration just after the instant of collapse, and
Figs. \skewpict (c)--(f) show the field evolution after the Hopf
texture collapse.

In their observational study of the collapse of Hopf textures in Nematic
liquid crystals, Chuang \etal\refmark\nlc
claimed that Hopf textures generally decay
by producing a global monopole-antimonopole pair. This
monopole-antimonopole pair is supposed to be produced at the point where
the Hopf texture first unwinds (on the negative $x$-axis in Fig. \skewpict).
The ``poles"  appear to be pulled in a roughly semicircular path (along
the circle, the generator of tori parameterized by $\eta$, in terms of
our convention) of until
they annihilate on the ``other side" of the Hopf texture (on the positive
$x$-axis in Fig. \skewpict).
NLC consists of rod-like molecules, and  the order parameter is
the orientation of the rod.
The rods do not carry senses ($\pm$), and hence flipping the rods
is a degeneracy operation. For example, would-be monopoles
and antimonopoles in $SO(3)$ field theory  are degenerate
in NLC. (Imagine the hedgehog configuration with each vector drawn without
an arrowhead.) The `equilibrium manifold' is $RP^2 = S^2/\pm$,
and the configuration in eq. \fieldII\ wraps
around the `equilibrium manifold' twice.
Since $RP^2$ can not be embedded in three space, there is no
half Hopf texture covering the `equilibrium manifold' once.
That being understood, we were curious about how the flow of
open `two spheres' can change into a flow of closed two spheres
converging into two parting monopoles.
Our simulations do show some resemblance to
this scenario in the gradient energy density,
but we do not observe the production of any
monopole-antimonopole pairs except on scale below our spacial resolution.
During the unwinding, the neighbouring fields in a cell rotate in the
opposite directions in our discrete sigma model, and this amounts to
fields' climbing over the potential barrier in the core in continuum
$\phi^4$ theory. Because of the rotation in the opposite directions,
there are cells which seem to contain monopoles during that brief
transition time, but they are not really monopoles with consistent
extension of the fields in the neighbouring cells outside the core.

This last point can also be seen by examining the output of two different
monopole detection algorithms: our standard monopole detection algorithm
(described in Ref. \BR), and another algorithm which simply searches for
points with $\vec\phi^2 \ll 1$. Clearly this later algorithm cannot be used
with the ``non-linear sigma model" simulations. For these simulations,
the monopole detection algorithm does ``detect" some monopoles, but
these have been determined to be numerical artifacts because their
location depends on the grid spacing but is nearly independent of all
other parameters such as the position in physical units. This diagnosis
is confirmed by the $\vec\phi^2 \ll 1$ detection algorithm for
some runs with finite $\lambda$. The maximum distance that these ``phantom
monopoles-antimonopole pairs" get from each other in any of these simulations
is only about $2\Delta x$ independent of the size of the initial Hopf
texture, so we are confident that they are not physical.
In our simulations, a pair (or pairs) of monopoles were not materialized
nor traveled along a `loop' over many cells.

\FIG\skewcoll{The time evolution of various fibers intersecting the $y=0$
plane is shown at time intervals of $1.6\Delta x$ from $t=0$ to just after
the collapse of the asymmetric Hopf
texture at $t\simeq 14\Delta x$. The solid lines and filled circles
represent the evolution of the $\phi_3=-1$ fiber, while the other curves
represent fibers in the $\vec\phi = (\cos\alpha,\sin\alpha,0)$ family
where $\alpha$ is an odd multiple of $\pi/8$.  }
Fig. \skewcoll\ shows the flow of a selected set of fibers during the collapse
of the asymmetric Hopf texture. The evolution of each side individually
resembles the collapse of the symmetric Hopf texture (Fig. \symcoll), but
there is a difference just after the collapse. For the symmetric Hopf texture,
many of the fibers disappear at the time of the collapse and do not ever
reappear (others reappear much later). All of the fibers shown in
Fig. \skewcoll\ crossing the $y=0$
plane on the ``small" ($x < 0$) side of the Hopf texture disappear at the
time of the Hopf texture collapse ($t = 14.4\Delta x$), but then they reappear
at $t=16.0\Delta x$ with the opposite orientation. This means that the
fibers have all converged at a point and jumped over the potential barrier
to ``untwist" the Hopf texture. In the configuration at $t=16.0\Delta x$
the fibers are arranged in a toroidal configuration as before, but now the
twist has been removed. This means that the same fibers are always
on the inside of the torus, so the inner fibers can shrink to a point and
disappear. This process proceeds with the outer fibers moving toward the
inside and then disappearing.

Finally, we have made some rather simple comparisons of the evolution of
the source term ($8\pi\dot{\vec\phi}^2$) for the growth of density fluctuations
in dark matter. We have compared the evolution of $8\pi\dot{\vec\phi}^2$ for
the collapsing texture simulations described above and for
simulations of the annihilation of a single monopole-antimonopole pair.
We found that although the two configuration had similar initial energies
and coherence scales, the collapsing Hopf texture collapsed more rapidly
and developed a higher kinetic energy by a factor of 2 or so. In addition,
collapsing Hopf textures also concentrated much of their energy into
a small central region during and after the collapse. These factors suggest
that the density fluctuations due to collapsing Hopf textures might be
more prominent than the density fluctuations due to monopole-antimonopole
annihilation. On the other hand, monopoles and antimonopoles are much
more numerous, so the most prominent annihilation induced fluctuations
might still be more prominent than the Hopf texture collapse induced
fluctuations. These questions can probably only be convincingly resolved
through by coupling gravity to the numerical simulations\rlap.
   \Ref\BRW{D. P. Bennett, S. H. Rhie, and D. Weinberg, in preparation, 1992.}

\chapter {Discussions}

This work started as the first step to find a way to measure
the relative population of
Hopf textures in comparison to that of monopoles in the simulations
of global monopole and texture seeded cosmic structure formation.\refmark\BR
Along the way, we became convinced that the Hopf textures are very
rare and the density fluctuation pattern due to Hopf textures are not
so dramatically different from that of monopoles. Perivolaropoulos
\Ref\PLEESE{L. Perivolaropoulos, Brown University preprint
            {\sl BROWN-HET}-775 (1990);
      R. Leese and T. Prokopec, Brown University preprint
            {\sl BROWN-HET}-778 (1990).}
has claimed that a $\pi_2$ texture of $SO(3)$ field theory
propagates with a planar structure, but the author used
 cylindrically symmetric
equations of motion ignoring the non-cylindrical modes.
In practice, the non-cylindrical modes prevent the `planar structure'
from being relevant.

Perhaps it is a different story if the
system is antiferromagnetic as in chiral spin liquid (CSL).
\Ref\CSLRF{ X. G. Wen {\it et al.}, {\sl Phys Rev.} {\bf B39} (1989) 11413;
          L. D. Laughlin and Z. Zou, {\sl Phys. Rev.} {\bf B41} (1989) 664;
          Stephen B. Libby, Z. Zou, and L. D. Laughlin  {\sl Nucl. Phys.}
          {\bf B348} (1991) 693.}
Laughlin, Zou, and Libby in Ref. \CSLRF\
claimed to have identified the gauge field of a non-abelian
magnetic monopole which is responsible for non-vanishing
Berry phases (for transport of spinons around a loop) in CSL.
P. B. Wiegmann recently
\Ref\WIEG {P. B. Wiegmann, {\it Nonabelian gauge theory of quantum
           antiferromagnetism in three dimensions and fractional quantum
           numbers of magnetic solitons}, IAS preprint (1991).}
argued that the `topological solitons' of CSL are $SO(3)$-magnetic
monopoles (Laughlin and his colleagues found to their distress
that their gauge fields do not take values in $so(3)$
but in a subspace of $su(4)$), and they must correspond to
`Hopf magnetic textures' (global `Hopf textures' in a magnetic system).
The implication of Wiegmann's claim that global monopoles can be identified
with Hopf textures is a bit puzzling. Nonetheless, it will be interesting
to know if $SO(3)$ topological
objects play roles in the dynamics of the antiferromagnetic system.
If we recall the work by Anderson
\Ref\ANDERSON {P. W. Anderson, {\sl Phys. Rev.} {\bf 86} (1952) 694.}
that the antiferromagnetic has a long-range order in the ground state,
where the ground state is a singlet state,
we can imagine a situation that  as the system cools, domains
of N\'eel state begin to form
such that the direction of the total spin of a sublattice in each
domain is random. Then the difference of the total spins of the
alternate sublattices resemble the isovector Higgs fields as an
order parameter. If they are relevant at all, one would like to know
what provides the stability of them ({\it e.g.,} textures collapse;
what forms the core of monopoles),
what the effects of doping would be, how they will translate into
the real situation
where the electromagnetic interaction is turned on, and so on.

\chapter{Conclusions}

We have explicitly constructed singly and multiply charged
Hopf texture configurations in
in an isovector scalar field theory which is spontaneously broken from
$SO(3)$ to $SO(2)$, and derived general formulas for the Hopf texture charge.
We have emphasized the importance of linking and gauge invariance.
We argued that partial Hopf textures are as rare as
full Hopf textures in the early universe, and made a crude estimation
that the  number density of horizon volume  should be smaller than
 $10^{-4}$.

By means of numerical simulations, we have followed the process of Hopf
texture collapse with both symmetric and asymmetric initial conditions.
We have shown that in neither of these cases does the Hopf texture
produce a monopole-antimonopole pair when it collapses. This is in contrast
to the observation by Chuang, \etal\ who claimed that the Hopf textures
in nematic liquid crystals decayed through the production of
global monopole-antimonopole pairs. It is possible that this discrepancy is
due to the differences in the topology or the dynamics between the
nematic liquid crystals and our $SO(3)$ scalar field theory. If this is
the reason, then it suggests that the nematic liquid crystals might be of
only limited use for the study of the evolution of topological defects in
a cosmological context.

Finally, we briefly discussed the role of Hopf textures in seeding density
fluctuations and found that although Hopf textures are much rarer than
annihilating monopole-antimonopole pairs, their density fluctuations maybe
somewhat more prominent.

\ack
This work was supported in part
the U.S. Department of Energy at the Lawrence Livermore
National Laboratory under contract No. W-7405-Eng-48
and by the NSF grant No. PHY-9109414.

\refout
\figout
\end